\RequirePackage[table]{xcolor}
\documentclass{PoS}
\usepackage{setspace}
\usepackage{slashed}
\usepackage{verbatim}    
\usepackage{graphicx} 
\usepackage{wrapfig}
\usepackage{dsfont}
\usepackage{epsfig}
\usepackage{epstopdf}
\usepackage{amsmath}
\usepackage{amsfonts,amssymb}
\usepackage{cleveref}
\usepackage{xcolor}

\newcommand{\be}{\begin{equation}}
\newcommand{\ee}{\end{equation}}
\newcommand{\bea}{\begin{eqnarray}} 
\newcommand{\eea}{\end{eqnarray}}
\newcommand{\MSbar}{{\overline{\rm MS}}}

\newcommand{\la}{\lambda}

\title{Renormalization of the Yukawa and Quartic Couplings in $\mathcal{N} = 1$ Supersymmetric QCD}
\ShortTitle{Yukawa and Quartic Couplings in Supersymmetric QCD}

    \author{M.~Costa$^{a,\, b}$, \speaker{H.~Herodotou}$^{ \,a}$, H.~Panagopoulos$^{\,a}$\\
	\llap{}$^a$Department of Physics, University of Cyprus, Nicosia, CY-1678, Cyprus\\
	$^b$Department of Chemical Engineering, Cyprus University of Technology, \\ 30 Archbishop Kyprianou Str., 3036, Limassol, Cyprus \\
	{\rm E-mail}:  \email{kosta.marios@ucy.ac.cy}, \email{herodotos.herodotou@ucy.ac.cy}, \email{panagopoulos.haris@ucy.ac.cy}}
	
\abstract{In this work we perform calculations in order to determine the renormalization factors and the mixing coefficients of the Yukawa and the quartic couplings in $\mathcal{N} = 1$ Supersymmetric QCD. The Yukawa couplings describe the interactions between gluino, quark and squark fields whereas the quartic couplings describe four-squark interactions. We discretize the action on a Euclidean lattice using the Wilson formulation for the gluino, quark and gluon fields; for squark fields (scalar fields) we employ na\"ive discretization. At the quantum level Yukawa and quartic interactions suffer from mixing with other operators which have the same transformation properties. Exploiting parity and charge conjugation symmetries of the Supersymmetric QCD action, we reduce the allowed mixing patterns. We compute, perturbatively to one-loop and to the lowest order in the lattice spacing, the relevant three-point Green’s functions so as to fine tune the Yukawa couplings and the relevant four-point Green’s functions to fine tune the quartic couplings. We use both dimensional and lattice regularizations as required for implementing the Modified Minimal Subtraction scheme ($\MSbar$). 
\begin{center}
\includegraphics[scale=0.45]{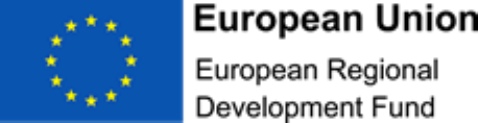}
\includegraphics[scale=0.45]{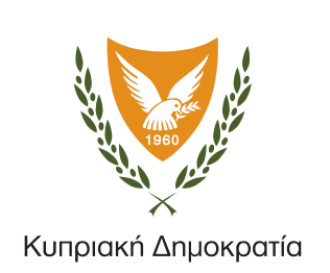}
\includegraphics[scale=0.45]{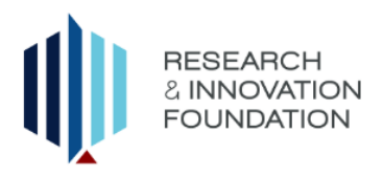}
\includegraphics[scale=0.45]{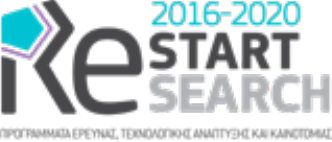}
\end{center}}

\FullConference{%
The 40th International Symposium on Lattice Field Theory,\\
31st July-4th August, 2023,\\
Fermilab, Batavia, Illinois, USA
}

\begin{document}
 
	\maketitle

	\section{Introduction}
\setcounter{page}{2}
Over the last decades, supersymmetry (SUSY) has been considered a prime candidate for resolving various open questions related to the Standard Model (SM), including the nature of dark matter and the unification of electromagnetic, weak, and strong forces proposed by the Grand Unified Theory (GUT). Supersymmetric theories of strongly interacting particles are actively explored to address these questions, with substantial efforts dedicated to examining the SUSY phase transitions and the SUSY breaking mechanism through numerical lattice simulations. Recently, there has been a growing focus on exploring supersymmetric QCD (SQCD) in this context. However, notable challenges arise due to the necessity of fine-tuning the bare parameters of the Lagrangian of the theory.


This work is a sequel to earlier investigations on SQCD and completes the one-loop fine-tuning of the SQCD action on the lattice, paving the way for numerical simulations of SQCD. More specifically, in Refs.~\cite{Costa:2017rht} and~\cite{Costa:2018mvb}, the first lattice perturbative computations in the context of SQCD were presented; apart from the Yukawa and quartic couplings \cite{Herodotou:2022xhz}, the renormalization of all parameters and fields appearing in the supersymmetric Lagrangian using Wilson gluons and fermions were extracted. Furthermore, the mixing of some composite operators under renormalization was explored. The results in these references will be employed in our current research.

In this study, we focus on lattice renormalization of the Yukawa and quartic couplings. Our approach involves employing the Wilson gauge action for gluon fields, the Wilson fermion action for fermions (quark and gluino fields), and a na\"ive discretization for squark fields. After introducing the basics of our computational setup (Section \ref{comsetUP}), we delve into a discussion of Yukawa coupling renormalization (Section \ref{couplingY}), both within dimensional and lattice regularization frameworks. We adopt the $\MSbar$ renormalization scheme and we determine the renormalization factors up to one-loop order. Similarly, we also provide preliminary results regarding the renormalization of quartic couplings (Section \ref{couplingQ}). To conclude, we end with a brief summary of our work and outline our future research plans (Section~\ref{summary}).
 
	\section{Formulation and Computational Setup}
	\label{comsetUP}

In this work we use the SQCD action in the Wess-Zumino (WZ) gauge \cite{Costa:2017rht, Costa:2018mvb, Curci:1987, Schaich:2014, Giedt:2014, Bergner:2016}. In our calculations, we use standard discretization; quarks ($\psi$), squarks ($A_\pm$) and gluinos ($\lambda$) are defined on the lattice points whereas gluons ($u_\mu$) are defined on the links between adjacent points: $U_\mu (x) = \text{exp}[i g a T^{\alpha} u_\mu^\alpha (x+a\hat{\mu}/2)]$; $\alpha$ is a color index in the adjoint representation of the gauge group. Below, we present the Euclidean action ${\cal S}^{L}_{\rm SQCD}$ on the lattice and in the massless limit using Wilson fermions:    
\begin{small}
\bea
{\cal S}^{L}_{\rm SQCD} & = & a^4 \sum_x \Big[ \frac{N_c}{g^2} \sum_{\mu,\,\nu}\left(1-\frac{1}{N_c}\, {\rm Tr} U_{\mu\,\nu} \right ) + \sum_{\mu} {\rm Tr} \left(\bar \lambda  \gamma_\mu {\cal{D}}_\mu\lambda  \right ) - a \frac{r}{2} {\rm Tr}\left(\bar \lambda   {\cal{D}}^2 \lambda  \right) \nonumber \\ 
&+&\sum_{\mu}\left( {\cal{D}}_\mu A_+^{\dagger}{\cal{D}}_\mu A_+ + {\cal{D}}_\mu A_- {\cal{D}}_\mu A_-^{\dagger}+ \bar \psi  \gamma_\mu {\cal{D}}_\mu \psi  \right) - a \frac{r}{2} \bar \psi   {\cal{D}}^2 \psi  \nonumber \\
&+&i \sqrt2 g_Y \big( A^{\dagger}_+ \bar{\lambda}^{\alpha}  T^{\alpha} P_+ \psi   -  \bar{\psi}  P_- \lambda^{\alpha}   T^{\alpha} A_+ +  A_- \bar{\lambda}^{\alpha}  T^{\alpha} P_- \psi   -  \bar{\psi}  P_+ \lambda^{\alpha}   T^{\alpha} A_-^{\dagger}\big)\nonumber\\  
&+& \frac{1}{2} g^2_4 (A^{\dagger}_+ T^{\alpha} A_+ -  A_- T^{\alpha} A^{\dagger}_-)^2 - m ( \bar \psi  \psi  - m A^{\dagger}_+ A_+  - m A_- A^{\dagger}_-)
\Big],
\label{susylagrLattice}
\eea
\end{small}where: $P_\pm= (1 \pm \,\gamma_5)/2$, $U_{\mu \nu}(x) =U_\mu(x)U_\nu(x+a\hat\mu)U^\dagger_\mu(x+a\hat\nu)U_\nu^\dagger(x)$, $a$ is the lattice spacing, $T^{\alpha}$ are the $SU(N_c)$ generators in the fundamental representation, $m$ is the mass of the matter fields (which may be flavor-dependent), $\cal{D}$ is the standard covariant derivative in the fundamental/adjoint representation \cite{Costa:2017rht}, $r$ is the Wilson parameter, $N_c$ is the number of colors, and a summation over flavors is understood in the last three lines of Eq.~(\ref{susylagrLattice}). Note that in the limit $a \to 0$ the lattice action reproduces the continuum one. To restore SUSY in the classical continuum limit, it is necessary for the values of $g_Y$ and $g_4$ at the tree level to coincide with the value of the gauge coupling, $g$.


Parity ($\cal{P}$) and charge conjugation  ($\cal{C}$) are symmetries of the continuum theory that is preserved exactly in the lattice formulation. The transformations of the fields under these symmetries are shown in Ref.~\cite{Herodotou:2022xhz}. Further symmetries of the classical action are: $\cal {R}$ [$U(1)_R$ rotation of the quark and gluino fields] and $\cal{\chi}$ [$U(1)_A$ axial rotation of the squark doublets ($A_+$, $A_-$) and the quark fields]. The $\cal{R}$ and $\cal{\chi}$ symmetries are broken on the lattice due to the Wilson terms.

	\section{Renormalization of the Yukawa Couplings}
	\label{couplingY}

To investigate the renormalization of the Yukawa couplings, we examine how dimension-4 operators, which are gauge-invariant, flavor singlets and contain a gluino, a quark, and a squark field, transform under the symmetries $\cal{P}$ and $\cal{C}$. Studying these operators, we conclude that there are two linear combinations of Yukawa-type operators which are invariant under $\cal{P}$ and $\cal{C}$ \cite{Giedt:2009}:
\bea
\label{chiInv}
Y_1 \equiv &&A^{\dagger}_+ \bar{\lambda} P_+ \psi   -  \bar{\psi}  P_- \lambda A_+ +  A_- \bar{\lambda} P_- \psi   -  \bar{\psi}  P_+ \lambda A_-^{\dagger} \\
Y_2 \equiv &&A^{\dagger}_+ \bar{\lambda} P_- \psi   -  \bar{\psi}  P_+ \lambda A_+ +  A_- \bar{\lambda} P_+ \psi   -  \bar{\psi}  P_- \lambda A_-^{\dagger} \, .
\label{chiNonInv}
\eea
Thus, all terms within each of the combinations in Eqs.~(\ref{chiInv}) and~(\ref{chiNonInv}) are multiplied by the same Yukawa coupling, $g_{Y_1}$ and $g_{Y_2}$, respectively. Note that the first combination corresponds to the third line in the Eq.~(\ref{susylagrLattice}). However, at the quantum level, the second (mirror) combination, may also emerge, with a Yukawa coupling, $g_{Y_2}$, that differs from the first one, $g_{Y_1}$. In the classical continuum limit, $g_{Y_1}$ equals to $g$, whereas $g_{Y_2}$ vanishes.

In a theory that includes massive quarks, ${\cal{R}}$ symmetry is no longer preserved. However, in the absence of anomalies, the ${\cal \chi} \times {\cal R}$ symmetry ensures that each component of the Yukawa term (as described in Eq.~(\ref{chiInv})) remains unchanged. On the other hand, it changes the components of the "mirror" Yukawa term (Eq.~(\ref{chiNonInv})) by introducing phase factors $e^{2i \theta}$ and $e^{-2i \theta}$.

\begin{figure}[ht!]
\centering
\includegraphics[scale=0.28]{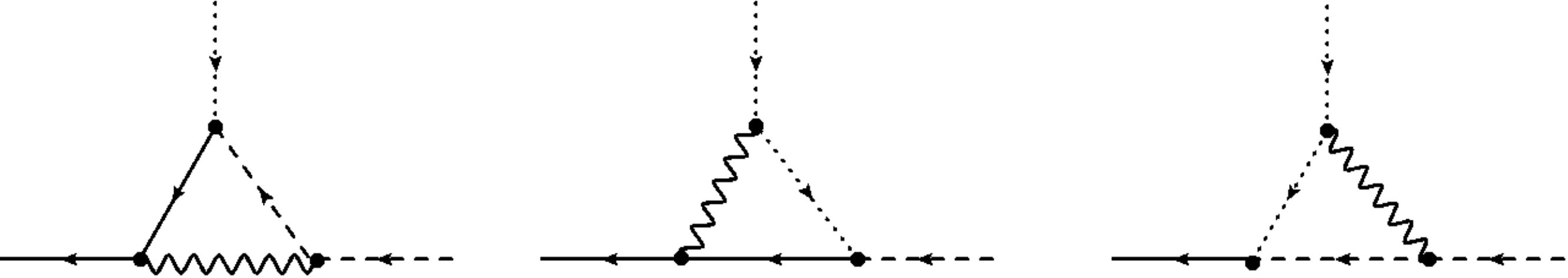}
\caption{One-loop Feynman diagrams leading to the fine-tuning of $g_{Y_1}$ and $g_{Y_2}$. A wavy (solid) line represents gluons (quarks). A dotted (dashed) line corresponds to squarks (gluinos). 
In the above diagrams the directions of the arrows on the external line depend on the particular Green's function under study. An arrow entering (exiting) a vertex denotes a $\la, \psi, A_+, A_-^{\dagger}$ ($\bar \la, \bar \psi, A_+^{\dagger}, A_-$) field. Squark lines could be further marked with a $+$($-$) sign, to denote an $A_+ \, (A_-)$ field.  
}
\label{couplingYukawa}
\end{figure}

In order to obtain the renormalization and mixing factors of the Yukuwa couplings, we compute, perturbatively, the relevant 3-point (3-pt) amputated Green's functions with external gluino-quark-squark fields using both dimensional regularization ($DR$) in $D = 4 - 2\epsilon$ dimensions and lattice regularization ($LR$). The three one-loop Feynman diagrams that enter the computation of these Green's functions appear in Fig. \ref{couplingYukawa}.  

Standard definitions of the renormalization factors of the fields and the gauge coupling constant $Z_g$ can be found, e.g., in Ref.~\cite{Costa:2017rht}. The Yukawa coupling is renormalized (at a reference scale $\mu$) as follows: 
\be
\label{gy}
g_{Y_1} \equiv g_{Y_1}^B = Z_{Y_1}^{-1} Z_g^{-1} \mu^\epsilon g^R,
\ee
where at the lowest perturbative order $Z_g Z_{Y_1} = 1$, and the renormalized Yukawa coupling $g_{Y_1}^R$ coincides with the gauge coupling $g^R$. We also define the renormalization factor for the squark fields as follows:
\be
\label{condS}
\left( {\begin{array}{c} A^R_+ \\ A^{R\,\dagger}_- \end{array} } \right)= \left(Z_A^{1/2}\right)\left( {\begin{array}{c} A^B_+ \\ A^{B\,\dagger}_- \end{array} } \right),
\ee
where $Z_A$ is a renormalization $2\times2$ mixing matrix. Within $DR$ regularization and $\MSbar$ renormalization this mixing matrix is diagonal~\cite{Costa:2017rht}. However, in a lattice regularization, the matrix is non-diagonal, and thus we have mixing between the components $A_+$ and $A_-^\dagger$.

We impose renormalization conditions which result in the cancellation of divergences in the corresponding bare 3-pt amputated Green's functions. Using the $\lambda-\bar\psi-A_+$ Green's function in $DR$, as an example, the renormalization condition can be expressed as follows: 
\be
\langle   \lambda(q_1)  {\bar{\psi}} (q_2) A_+(q_3) \rangle \Big \vert^\MSbar \equiv Z_\psi^{-1/2} Z_\la^{-1/2}  (Z_A^{-1/2})_{++} \langle   \lambda(q_1)  {\bar{\psi}} (q_2) A_+(q_3) \rangle \Big \vert^{\rm{bare}} \, .
\label{renormC}
\ee
Coupling constants appearing in the right-hand side of Eq.~(\ref{renormC}) must be expressed in terms of their renormalized values; doing so will involve use of Eq.~(\ref{gy}), and thus will lead to a determination of $Z_Y$. The left-hand side of Eq.~(\ref{renormC}) is just the $\MSbar$ (free of pole parts) renormalized Green's function. The other renormalization conditions which involve squark fields $A_+^\dagger, \, A_-, \, A_-^\dagger$ are variants of Eq.~(\ref{renormC}).

The calculations in this work could ideally be carried out using arbitrary external momenta. However, for convenience of computation, after confirming that there are no infrared (IR) divergences, we proceed with the computation of the relevant diagrams by setting one of the external momenta to zero. Using the values of the renormalization factors of the fields and of the gauge coupling in $DR$ \cite{Costa:2017rht}, all variants of Eq.~(\ref{renormC}) lead to an identical value for $Z_Y^{DR, \MSbar}$:
\be
Z_Y^{DR, \MSbar}= 1 + {\cal O}(g^4) .
\label{ZYDR}
\ee
Therefore, we conclude that, at the quantum-level, the renormalization of the Yukawa coupling in DR is not affected by one-loop corrections. We also expect that the corresponding renormalization on the lattice will be finite.

As previously stated, on the lattice, the renormalization matrix $Z_A$ is non-diagonal; mixing between the squark components appears on the lattice through the finite nondiagonal elements of $Z_A$. In addition, as a result of utilizing Wilson discretization, the ${\cal \chi} \times {\cal R}$ symmetry is broken, causing lattice bare Green's functions not to be invariant under ${\cal \chi} \times {\cal R}$ at the quantum level. Therefore, in the calculation of the lattice bare Green's functions, one-loop spurious contributions will arise, which will need to be removed by introducing the mirror Yukawa counterterm in the action. Taking as an example the Green's function with squark $A_+$, the renormalization condition on the lattice reads: 
\begin{small}
\be
\langle   \lambda(q_1)  {\bar{\psi}} (q_2) A_+(q_3) \rangle \Big \vert^\MSbar \equiv Z_\psi^{-1/2} Z_\la^{-1/2}  \langle   \lambda(q_1)  {\bar{\psi}} (q_2) \bigl((Z_A^{-1/2})_{++} A_+(q_3) + (Z_A^{-1/2})_{+-} A_-^\dagger(q_3)\bigr) \rangle \Big \vert^{\rm{bare}} \, .
\label{renormCLatt}
\ee
\end{small} 

The bare couplings on the right-hand side of this equation must be expressed in terms of the corresponding renormalized ones by using $Z_g$ and $Z_{Y_1}$; a mirror Yukawa term also contributes, with a coupling constant $g_{Y_2}$ which will be determined in what follows. Eq. (\ref{renormCLatt}) consists of two types of contributions with opposite chiralities. Consequently, two separate conditions will be used to determine the two unknowns $Z_{Y_1}$ and $g_{Y_2}$. 

From Eq.~(\ref{renormCLatt}) at first perturbative order, ${\cal O}(g^2)\,$, the difference between the one-loop $\MSbar$-renormalized Green's functions and the corresponding lattice bare Green's functions allows us to deduce $Z_{Y_1}$ and $g_{Y_2}$. By combining this difference and the renormalization factors of fields and gauge coupling on the lattice, we obtain the renormalization factors: 
\begin{align}
        {Z_{Y_1}}^{LR,\MSbar} &= 1 +  \frac{g^2}{16\,\pi^2}\bigg(\frac{1.45833}{N_c} + (4.40768- 2 c_{\rm hv})N_c + 0.520616 N_f \bigg)\\
        {g_{Y_2}}^{LR,\MSbar} &=\frac{g^3}{16\,\pi^2}\bigg(\frac{-0.040580}{N_c} + (2.45134 - 2 c_{\rm hv} )N_c \bigg) ,
\end{align}
where  $c_{\rm hv} = 0 \, (1)$ for the na\"ive ('t Hooft-Veltman (HV)) prescription of $\gamma_5$; while the choice of $c_{\rm hv}$ cannot affect pole parts, it is present in finite contributions, as seen above. 
As expected from general renormalization theorems, the $\MSbar$ renormalization factors for gauge invariant objects are gauge-independent, as in this case. Furthermore, the multiplicative renormalization $Z_{Y_1}$ and the coefficient $g_{Y_2}$ of the mirror Yukawa counterterm are finite as we can predict from the continuum calculation. 

\section{Renormalization of the Quartic Couplings}
\label{couplingQ}

\begin{figure}[ht!]
\centering
\includegraphics[scale=0.34]{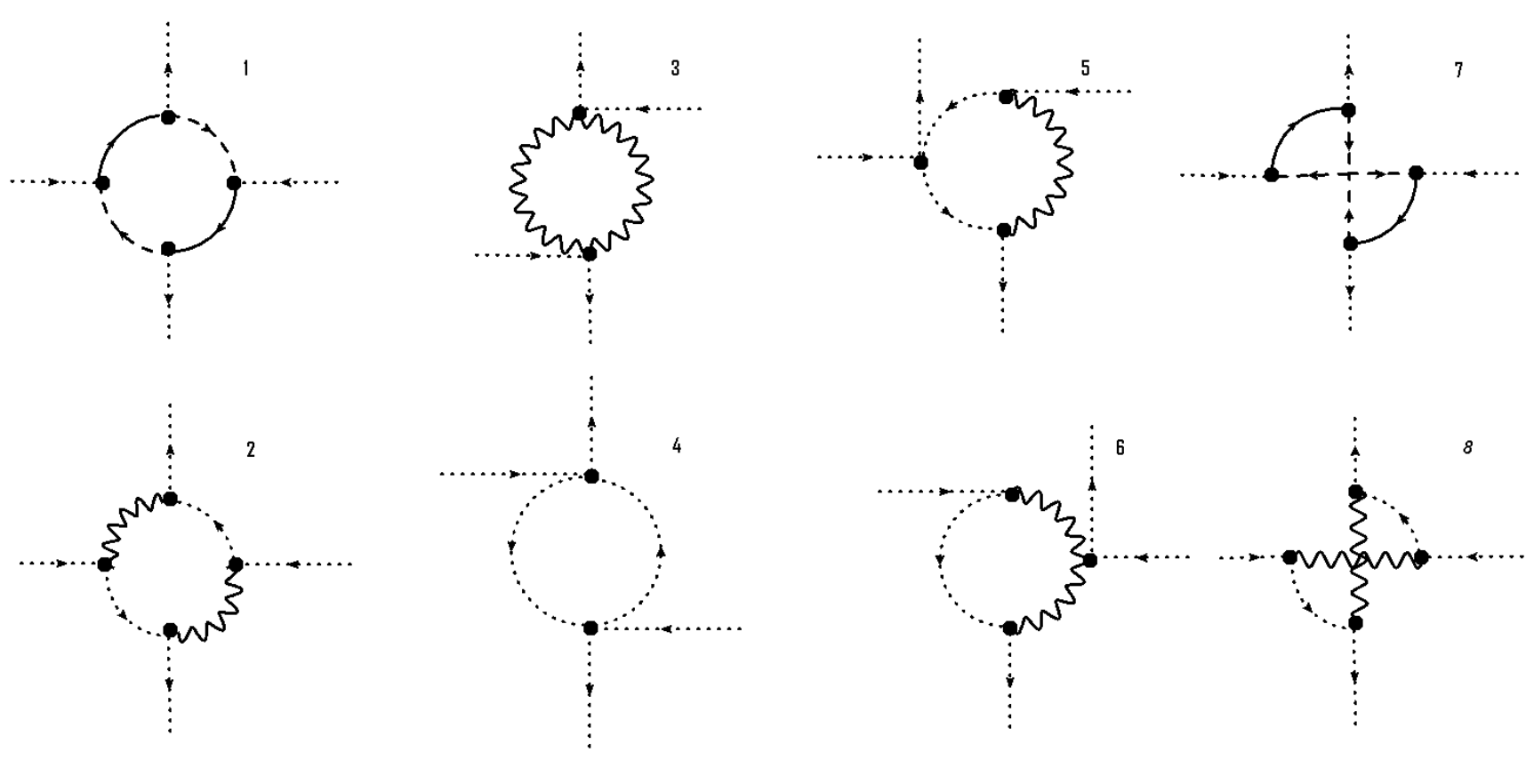}
\includegraphics[scale=0.34]{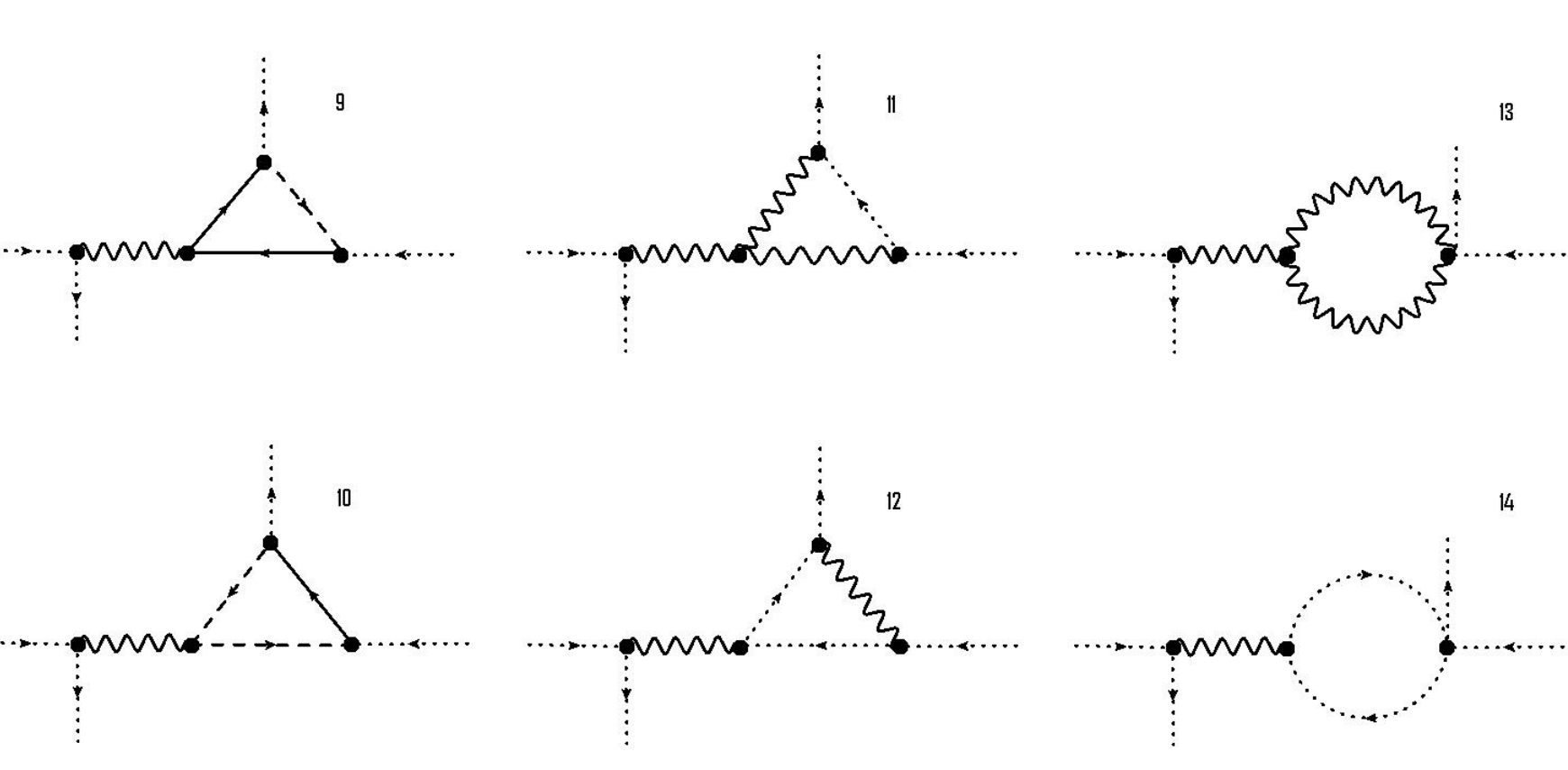}
\includegraphics[scale=0.33]{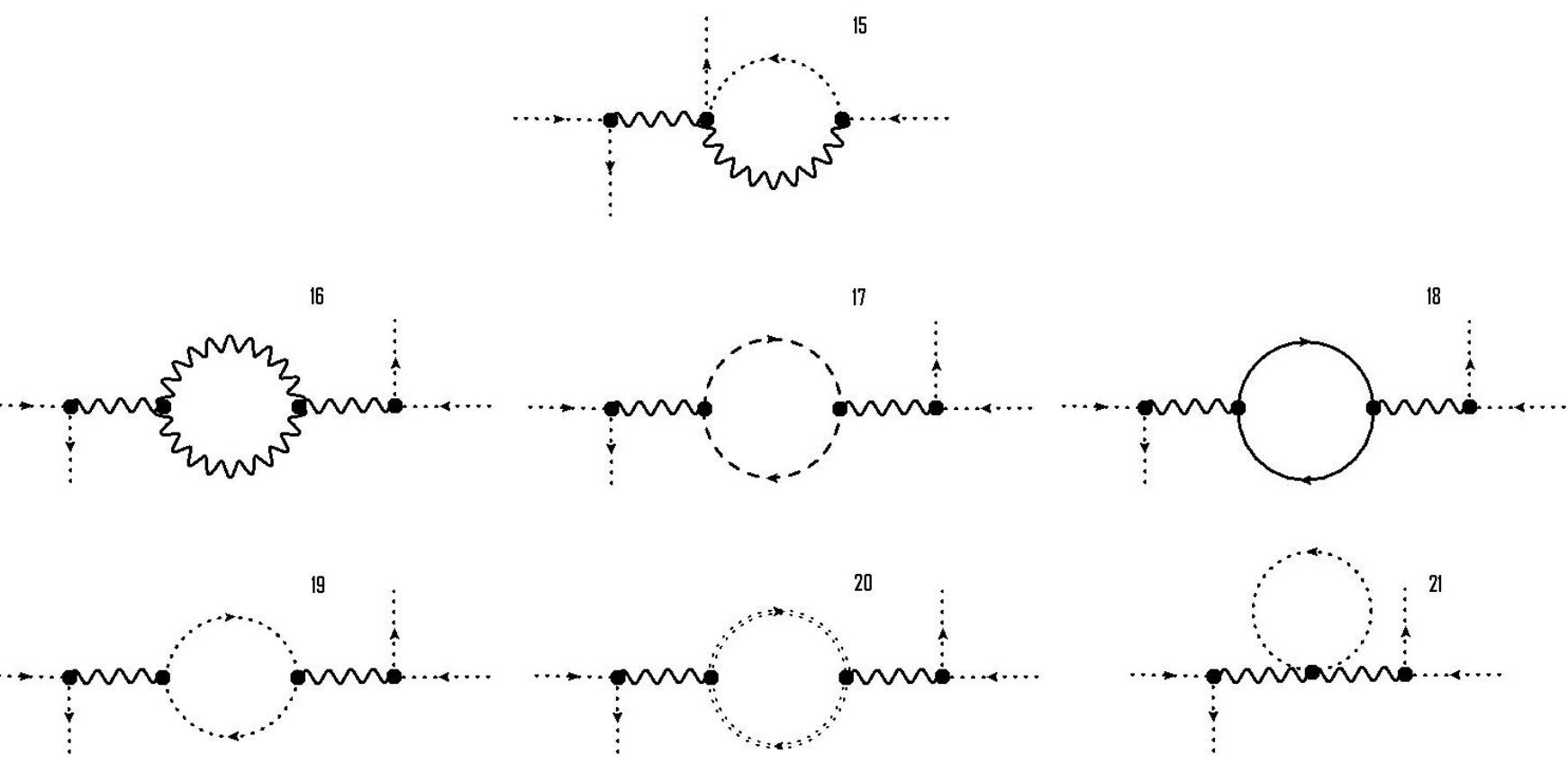}
\caption{One-loop Feynman diagrams leading to the fine-tuning of $\la_i$.  Notation is identical to that of Fig.~\ref{couplingYukawa}. A double-dotted line stands for ghost fields.
}
\label{couplingquartic}
\end{figure}

In order to determine the renormalization factors of the quartic couplings (four-squark interactions), we have to calculate Green's functions with four squark fields; two of them have to lie in the fundamental representation and the other two in the antifundamental. There are ten cases for choosing these squarks. We have to construct combinations that remain unchanged under the symmetries of $\cal{C}$ and $\cal{P}$. Table~\ref{tb:nonsinglet2} displays all of these combinations \cite{Wellegehausen:2018opt}. 
The tree-level values of the quartic couplings, $\la_i$, shown in Table~\ref{tb:nonsinglet2}, are:
\be
\la_1 =  g^2,  \, \, \, \la_2 = \la_3 = \la_4= \la_5 =0 \, .
\ee
These couplings receive quantum corrections, coming from the Feynman diagrams of Fig.~\ref{couplingquartic}. The first eight Feynman diagrams (along with various mirror versions) are 1PI, the rest of them are not. These one-loop diagrams enter the computation of the 4-point amputated Green's functions for the quartic couplings. We compute, perturbatively, the relevant Green's functions using both dimensional and lattice regularization. The Majorana nature of gluinos shows up in diagrams 7 and 10, in which $\lambda-\lambda$ as well as $\bar{\lambda}-\bar{\lambda}$ propagators appear. 

\clearpage

\begin{table}[ht!]
\begin{center}
\scalebox{0.8}{
  \begin{tabular}{c|c|c}
\hline \hline
Operators & $\cal{C}$&$\cal{P}$ \\ [0.5ex] \hline\hline
    $\la_1 (A_+^{\dagger} \, T^{\alpha} \, A_+ + A_- \, T^{\alpha} \, A_-^{\dagger})^2/2 $&$+$& $+$   \\[0.5ex]\hline
    $\la_2 [(A_+^{\dagger}  A_-^{\dagger})^2 + (A_- A_+)^2] $&$+$& $+$   \\[0.5ex]\hline
    $\la_3  (A_+^{\dagger}  A_+)  (A_- A_-^{\dagger}) $&$+$ & $+$  \\[0.5ex]\hline
    $\la_4 (A_+^{\dagger}  A_-^{\dagger}) (A_- A_+) $&$+$& $+$   \\[0.5ex]\hline
    $\la_5 (A_+^{\dagger}  A_-^{\dagger} + A_- A_+)(A_+^{\dagger}  A_+  + A_- A_-^{\dagger}) $&$+$ & $+$  \\[0.5ex]\hline
\hline
\end{tabular}}
\caption{Dimension-4 operators which are gauge invariant and flavor singlets. All operators appearing in this table are eigenstates of charge conjugation, $\cal{C}$, and parity, $\cal{P}$, with eigenvalue 1. In the above operators, squark fields carry flavor indices. The symbols $\la_i$ are five quartic couplings.}
\label{tb:nonsinglet2}
\end{center}
\end{table}

\noindent There are also additional one-loop Feynman diagrams leading to the fine-tuning of the quartic couplings on the lattice.


The computation of all 4-pt diagrams is currently in progress. Our results are presented in two papers: one concerns the Yukawa couplings (gluino-quark-squark interactions) \cite{Yukawa:2023}, and the other, forthcoming, focuses on the quartic couplings (4-squark interactions) in SQCD \cite{quartic:2023}.

\section{Summary -- Future Plans}
\label{summary}

In this research project, we investigate the renormalization factors of the Yukawa and quartic couplings in the framework of $\mathcal{N} = 1$ Supersymmetric QCD. In order to determine these factors, we compute, perturbatively, 3-pt and 4-pt Green's functions using both dimensional and lattice regularizations. We conclude that the renormalization factor of the Yukawa coupling and the coefficient of the mirror Yukawa counterterm are finite on the lattice. In our ongoing investigation we are calculating perturbatively the relevant four-point Green’s functions so as to deduce the renormalization of the quartic couplings. Applying these results to the bare couplings in Monte Carlo simulations of SQCD would be a first step in the non-perturbative fine-tuning of the lattice action. An extension of our work would be the perturbative investigation of Supersymmetric non-abelian models on the lattice through the use of chirally invariant actions.

{\bf Acknowledgements:} 
 This work was co-funded by the European Regional Development Fund and the Republic of Cyprus through the Research and Innovation Foundation (Projects: EXCELLENCE/0421/0025). M.C. also acknowledges partial support from the Cyprus University of Technology under the ``POST-DOCTORAL" programme.


\begin{thebibliography}{99}
		\bibitem{Costa:2017rht}
		M.~Costa and H.~Panagopoulos,
		Phys. Lett. D \textbf{76} (2017), 094514
		[arXiv:1706.05222].
		
		\bibitem{Costa:2018mvb}
		M.~Costa and H.~Panagopoulos,
		Phys. Lett. D \textbf{99} (2019), 074512
	    [arXiv:1812.06770].
     
        \bibitem{Herodotou:2022xhz}
         M.~Costa, H.~Herodotou and H.~Panagopoulos,
        PoS \textbf{LATTICE2022} (2023), 276
        doi:10.22323/1.430.0276
        [arXiv:2210.03695 [hep-lat]].
        
        \bibitem{Curci:1987}
		G. Curci and G. Veneziano,
		Nucl. Phys. B \textbf{292} (1987), 555.

       	\bibitem{Schaich:2014}
		D. Schaich,
		PoS (\textbf{LATTICE2018}) 005
        [arXiv:1810.09282].
 
		\bibitem{Giedt:2014}
		S. Catterall, D. Schaich, P.H. Damgaard, T. DeGrand and J.~Giedt,
		Phys. Rev. D \textbf{90} (2014), 065013
        [arXiv:1405.0644].
		
		\bibitem{Bergner:2016}
		S. Catterall and G. Bergner,
		Mod. Phys. A \textbf{31} (2016), 1643005 
		[arXiv:1603.04478].
  
		\bibitem{Giedt:2009}
		J.~Giedt,
        Mod. Phys. A \textbf{24} (2009), 4045-4095
        [arXiv:0903.2443].
		
		\bibitem{Wellegehausen:2018opt}
		B.~Wellegehausen and A.~Wipf,
        PoS (\textbf{LATTICE2018}) 210
		[arXiv:1811.01784].

		
  
        \bibitem{Yukawa:2023}
        M.~Costa and H.~Herodotou,
        (2023), [arXiv:2310.202071].

        \bibitem{quartic:2023}
        M.~Costa, H.~Herodotou and H.~Panagopoulos,
        ``Supersymmetric QCD on the Lattice: Fine-Tuning and Counterterms for the Quartic Couplings'', 
        in preparation.
		
	\end{thebibliography}
\end{document}